\documentclass[noshowpacs,preprintnumbers,secnumarabic,amsmath,amssymb,nofootinbib]{revtex4}

\usepackage{graphicx}
\usepackage{dcolumn}
\usepackage{bm}

\def\beq{\begin{equation}}
\def\eeq{\end{equation}}
\def\be{\begin{equation}}
\def\ee{\end{equation}}
\def\bea{\begin{eqnarray}}
\def\eea{\end{eqnarray}}
\def\d{{\rm d}}
\def\cH{{\cal H}}

\def\tk{\tilde{k}}
\def\tK{\tilde{K}}
\def\vk{\mathbf{k}}
\def\vtk{\mathbf{\tilde k}}
\def\vK{\mathbf{K}}
\def\vtK{\mathbf{\tilde K}}

\DeclareRobustCommand{\SkipTocEntry}[4]{}

\begin{document}

\vspace{5cm}
\title{Gravitational Wave Spectrum Induced by Primordial Scalar Perturbations}

\author{Daniel Baumann$^1$}
\email{dbaumann@princeton.edu}

\author{Paul Steinhardt$^{1,2}$}
 \email{steinh@princeton.edu}

\author{Keitaro Takahashi$^1$}
 \email{ktakahas@princeton.edu}

\affiliation{%
$^1$Department of Physics, Jadwin Hall, Princeton University, 
Princeton, NJ 08544\\
$^2$Princeton Center for Theoretical Physics, Jadwin Hall, \\
Princeton University, Princeton, NJ 08544}

\author{Kiyotomo Ichiki}
\email{ichiki@resceu.s.u-tokyo.ac.jp}

\affiliation{Research Center for the Early Universe, University of Tokyo,\\
7-3-1 Hongo, Bunkyo-ku, Tokyo 113-0033, Japan}


\begin{abstract}
We derive the complete spectrum of gravitational waves induced by primordial scalar perturbations ranging over all observable wavelengths.  This scalar-induced contribution can be computed directly from the observed scalar perturbations and general relativity and is, in this sense, independent of the cosmological model for generating the perturbations. 
The spectrum is scale-invariant on small scales, but has an interesting scale-dependence on large and intermediate scales, where scalar-induced gravitational waves do {\it not} redshift and are 
hence enhanced relative to the background density of the Universe. 
This contribution to the tensor spectrum is significantly different in form from the direct model-dependent primordial tensor spectrum and, although small in magnitude, it dominates the primordial signal for some cosmological models.
We confirm our analytical results by direct numerical integration of the equations of motion.
\end{abstract}



\maketitle
\tableofcontents


\section{Introduction}
\label{sec:intro}

Arguably the most striking prediction of inflationary cosmology \cite{Inflation} 
is the causal generation of 
nearly scale-invariant spectra of 
both scalar (energy density) and tensor (gravitational wave, GW) perturbations. 
The natural prediction is that the scalar and tensor amplitudes are comparable 
within one or two orders of magnitude of one another by virtue of the fact that 
both are created by the same de Sitter quantum process.
The existence of a scalar spectrum is now firmly 
established by measurements of the cosmic microwave background (CMB) \cite{WMAP}
and large scale structure \cite{SDSS}, and its amplitude is well-determined.   
Tensor 
fluctuations, on the other hand, have yet to be detected, although current 
measurements 
have only begun to probe the expected range of amplitudes.  

Detecting primordial tensor fluctuations is an important milestone because it 
rules out a 
whole class of alternative cosmological scenarios, like the ekpyrotic 
\cite{Ekpyrotic} and cyclic models \cite{Cyclic}, 
which produce virtually the identical scalar spectrum as inflation but a 
completely 
different tensor spectrum.  In particular, the primordial tensor contribution in 
ekpyrotic/cyclic models is exponentially smaller and more blue 
\cite{LathamCyclic}.  
Detection of a primordial tensor signal is therefore widely regarded as a 
smoking gun signature of inflation.
However, failing to detect the tensor modes at the expected level does not 
necessarily rule out 
inflation.  The inflationary tensor signal can be suppressed by extra fine-tuning of 
the inflationary model and/or the addition of extra fields (e.g. hybrid 
inflation \cite{hybrid}) so 
that the background equation of state of the Universe, instead of changing 
smoothly during the final stages of 
inflation, undergoes a sequence of jerks and gyrations \cite{Latham}.   
A number of studies have discussed 
the limits to how far a search for the tensor spectrum can go based on detector 
sensitivity 
and foregrounds \cite{CMB}.

\begin{figure}[h] 
   \centering
   \includegraphics[width=0.65\textwidth]{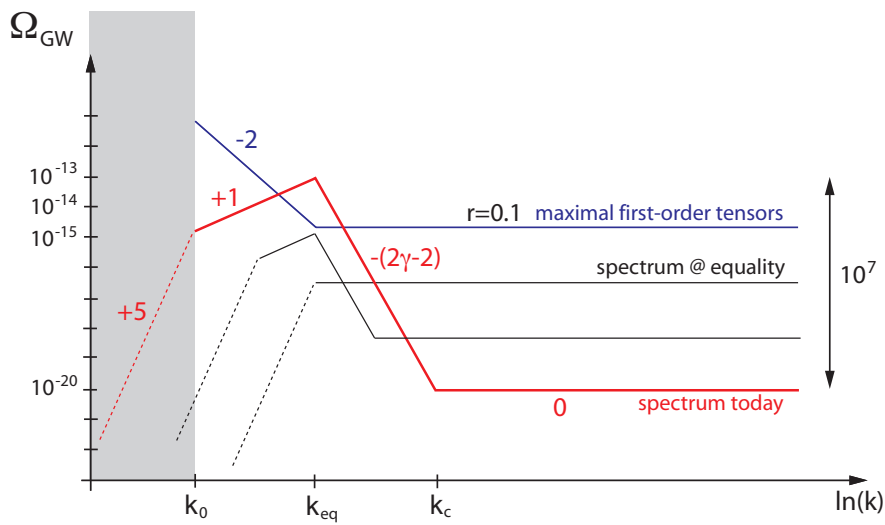} 
   \caption{{\bf Spectra of first- and second-order gravitational waves:} This 
schematic 
illustrates the conjectured form of $\Omega_{\rm GW}(k)$, the fraction of the 
critical density in  gravitational 
waves per log-interval of wavenumber $k$, as
derived in section \ref{sec:analytic}. The topmost curve represents the typical 
first-order 
inflationary tensor spectrum.  With fine-tuning, it can  be suppressed below the 
level of 
the second-order, scalar-induced tensor perturbations (bottom curves).    The 
bottom curves 
represent a sequence of times:  matter-radiation equality ($a_{\rm 
eq}$), redshift $z=100$, and today ($a_0$). 
The scalar-induced tensor spectra shown here are for a perfectly scale-invariant scalar input spectrum ($n_s = 1$).
If the scalar spectrum is blue ($n_s > 1$) the induced tensor spectrum is enhanced on small scales (large $k$), while a red spectrum ($n_s < 1$) suppresses tensor fluctuations on small scales (see section \ref{sec:discussion} for cautionary remarks about extrapolating spectra to very small scales using the large-scale power law form of the scalar spectrum).
$\Omega_{\rm GW}$ is of course ill-defined on superhorizon scales. On superhorizon scales (dashed lines) we therefore formally define the rescaled tensor power spectrum, $k^2 P_h(k)$, but do {\it not} interpret it as an energy density of gravitational waves (see section \ref{sec:analytic}). }
   \label{fig:1}
\end{figure}

At second order in perturbation theory the observed scalar spectrum sources the generation of secondary tensor modes \cite{GW2}.
In this paper, we analyze the stochastic spectrum of second-order gravitational 
waves 
induced by the first-order scalar perturbations.  Since the scalar 
spectrum is already 
measured, this contribution to the tensor spectrum must exist and must be the 
same for 
both inflationary and ekpyrotic models because their predictions for the scalar 
spectrum 
match.  For inflation, this second-order contribution is generically negligible, 
orders of 
magnitude smaller than the first-order contribution except for models with 
extreme fine-tuning.  For ekpyrotic and cyclic models, the scalar-induced 
second-order contribution 
computed here is actually the dominant contribution on astrophysical and 
cosmological 
scales, because the first-order tensor spectrum is always exponentially small
compared to the scalar spectrum.  
Hence, the calculation here supersedes previous predictions of the tensor 
spectrum for 
ekpyrotic and cyclic models \cite{LathamCyclic}.  
Because the gravitational wave spectrum we compute here is purely a consequence 
of the {\it observed} scalar spectrum and general relativistic evolution, any 
mechanism that accounts for the observed spectrum of scalar fluctuations also 
generates the same secondary tensor spectrum, provided Einstein's equations 
hold. Hence, this second-order signal provides an absolute {\it lower limit} on 
tensors from the early Universe.\\

Our work builds on important earlier work by Mollerach, Harari and Matarrese 
\cite{Mollerach} 
and Ananda, Clarkson and Wands \cite{Wands}.    Mollerach {\it et al.} 
\cite{Mollerach} computed the 
effect of second-order gravitational waves on large-scale CMB polarization. They 
found that the second-order 
tensors dominate over the first-order signal if the ratio of the tensor-to-scalar amplitude 
on the largest observable wavelengths is $r < 10^{-6}$. 
Then, more recently, Ananda, Clarkson and Wands \cite{Wands} numerically studied 
the 
present spectrum of gravitational waves on very small scales accessible to 
direct detection experiments like the Big Bang Observer (BBO). Typically, the 
signal is expected to be at the extreme limit of the predicted sensitivity of 
BBO.
Here we compute the complete spectrum of scalar-induced gravitational 
waves on all scales and discuss how it evolves with time. We analytically 
determine a 
critical scale above which 
second-order gravitational waves do {\it not} redshift. This non-trivial 
transfer function for second-order gravitational waves leads to an interesting 
feature in the 
current spectrum (see the schematic in Figure \ref{fig:1}) with a factor $10^7$ 
enhancement of modes of order the horizon size at matter-radiation equality 
relative to the scale-invariant small scale spectrum. We confirm our analytical 
findings by numerical integration of the equations of motion.\\

The outline of the paper is as follows:
In \S \ref{sec:2t}, we derive the evolution equations for second-order tensor 
fluctuations sourced by first-order scalar fluctuations. Allowing for an  
anisotropic stress contribution to the energy momentum tensor, we derive a 
general expression for the power spectrum of scalar-induced gravitational waves. 
This generalizes the work of Ref.~\cite{Wands}. 
In \S\ref{sec:analytic}, we analyze the spectrum using various approximations 
and 
scaling arguments.  These analytical estimates are confirmed  by direct 
numerical 
integration of the equations of motion in \S\ref{sec:numeric}. Finally, we 
discuss the 
implications of these results in 
\S\ref{sec:discussion}. 
In two appendices we recall the Green's functions for gravitational waves and the transfer functions for first-order scalar fluctuations \cite{Mukhanov}. \\

We use the following conventions: Throughout we employ natural units, $\hbar = c \equiv 1$, and (reduced) Planck mass
$M_P^{-2} = 8 \pi G \equiv \kappa^2$, as well as 'East coast' signature for the metric, 
$(-,+,+,+)$. Greek indices, $\mu, \nu = 0, \dots, 3$, denote four-dimensional 
spacetime indices, while roman indices, $i, j=1, \dots, 3$, are reserved for 
spatial indices. The parameter $\eta$ is conformal time, $a(\eta)\, \d \eta = \d 
t$.

\section{Second-Order Tensors from First-Order Scalars}
\label{sec:2t}

Let us recall some basic facts about second-order perturbation theory, before 
deriving the explicit form of the evolution equations for second-order, scalar-induced tensors.
We consider perturbations to a flat Friedmann-Robertson-Walker (FRW) background, 
$g^{(0)}_{\mu \nu} = a^2(\eta) \eta_{\mu \nu}$, 
\beq
g_{\mu \nu} = g_{\mu \nu}^{(0)} + \delta g_{\mu \nu}\, ,
\eeq
where $g_{\mu \nu}^{(0)}$
satisfies the 0th-order Einstein equations, $G^{(0)}_{\mu \nu} = \kappa^2 
T^{(0)}_{\mu \nu}$,
\beq
\cH^2 = \frac{\kappa^2 a^2}{3} \rho^{(0)}\, ,
\quad \cH^2- \cH'  = \frac{\kappa^2 a^2}{2} (\rho^{(0)} + P^{(0)})\, ,
\quad \cH \equiv \partial_\eta \ln a\, .
\eeq
Here $\rho^{(0)}$ and $P^{(0)}$ are the homogeneous background density and pressure, respectively, and $(\dots)'$ denotes a derivative with respect to conformal time, $\eta$.
Including to linear order the small quantum perturbations to the metric and energy density, the solution to the 1st-order Einstein equations,
$G^{(1)}_{\mu \nu} = \kappa^2 T^{(1)}_{\mu \nu}$,  can be decomposed into independent scalar, vector and tensor modes.
At linear order, different $k$-modes in Fourier space are independent. This is in contrast to the 
2nd-order Einstein equations,
$G^{(2)}_{\mu \nu} = \kappa^2 T^{(2)}_{\mu \nu}$, where different $k$-modes mix and scalar,
vector and tensor modes are not independent. 
However, it is important to notice that, at second order, there is no mixing between second-order scalar, vector, and tensor modes.  On the other hand, there is a second-order contribution to the tensor mode, $h_{ij}^{(2)}$, that depends quadratically on the first-order scalar metric perturbation.  
This contribution, the "scalar-induced" tensor mode, is the focus of this paper.\footnote{Scalar-induced vector modes were studied in \cite{Vectors}.}

\subsection{Evolution Equations}

To compute the second-order, scalar-induced tensor mode we begin with the following perturbed metric
\beq
\label{equ:metric}
\d s^2 = a^2(\eta) \left[-\left(1 + 2 \Phi^{(1)} + 2 \Phi^{(2)} \right) \d 
\eta^2 + 2 V^{(2)}_i \d \eta \d x^i
       + \left\{ \left(1 - 2 \Psi^{(1)} - 2 \Psi^{(2)} \right) \delta_{ij} + 
\frac{1}{2} h_{ij} \right\} \d x^i \d x^j \right]\, ,
\eeq
where 
$h_{ij} \equiv h^{(2)}_{ij}$ and we have ignored first-order vector and tensor 
perturbations. 
Here and in the following, the superscripts are formal labels for the order of the perturbation.
The second-order Einstein tensor
and energy-momentum tensor are \cite{Acquaviva}
\bea
G^{(2)i}_{~~~~j}
&=& a^{-2} \Biggl[ \frac{1}{4} \left( h^{i}_{j}{}'' + 2 \cH h^{i}_{j}{}' - 
\nabla^2 h^{i}_{j} \right)
    + 2 \Phi^{(1)} \partial^i \partial_j \Phi^{(1)} - 2 \Psi^{(1)} \partial^i 
\partial_j \Phi^{(1)}
    + 4 \Psi^{(1)} \partial^i \partial_j \Psi^{(1)} \nonumber \\
& & ~~~~~~
    + \partial^i \Phi^{(1)} \partial_j \Phi^{(1)} - \partial^i \Phi^{(1)} 
\partial_j \Psi^{(1)}
    - \partial^i \Psi^{(1)} \partial_j \Phi^{(1)} + 3 \partial^i \Psi^{(1)} 
\partial_j \Psi^{(1)}
    \nonumber \\
& & ~~~~~~
    + \left( \Phi^{(2)}, \Psi^{(2)}, V^{(2)}_i ~ {\rm terms} \right)
    + \left({\rm diagonal \, \, part} \right) \delta^{i}_{j}  \Biggr],
\eea
and
\beq
T^{(2)i}_{~~~~j}
= \left( \rho^{(0)} + P^{(0)} \right) v^{(1)i} v^{(1)}_j + P^{(0)} 
\Pi^{(2)i}_{~~~~j}
  + P^{(1)} \Pi^{(1)i}_{~~~~j} + P^{(2)} \delta^{i}_{j}\, ,
\eeq
where $\rho$, $P$, $v$ and $\Pi$ are energy density, pressure, velocity and anisotropic stress, respectively.
We act on the spatial components of the Einstein equations with the projection 
tensor $\hat{\cal T}_{ij}^{~lm}$ \cite{Wands},
\beq
\label{equ:E2}
\hat{\cal T}_{ij}^{~lm} G^{(2)}_{lm} = \kappa^2 \hat{\cal T}_{ij}^{~lm} 
T^{(2)}_{lm}\, .
\eeq
We will define the operator $\hat{\cal T}_{ij}^{~lm}$ explicitly below, but we note here 
that it extracts the transverse, traceless part of any tensor and
eliminates the terms involving $\Phi^{(2)}, \Psi^{(2)}, V^{(2)}_i, P^{(2)}$ and
the scalar and vector parts of $\Pi^{(2)i}_{~~~~j}$ in the second-order Einstein 
equations.
Using the following first-order relations,
\bea
&&
P^{(1)} \equiv c_{s}^{2} \rho^{(1)} \label{equ:cs} \, ,
\\ 
&&
\rho^{(1)}
        = - \frac{2 }{\kappa^2 a^2}
            \left[ 3 \cH \left(\cH \Phi^{(1)} - \Psi^{(1)}{}' \right) + \nabla^2 
\Psi^{(1)} \right],
\\ 
&&
v^{(1)}_i = - \frac{2}{\kappa^2 a^2 (\rho^{(0)} + P^{(0)})}
            \partial_i \left( \Psi^{(1)}{}' + \cH \Phi^{(1)} \right),
\\ 
&&
\Pi^{(1)i}_{~~~~j}
= - \frac{1}{\kappa^2 a^2 P^{(0)}}
    \left( \partial^i \partial_j - \frac{1}{3} \delta^{i}_{j} \nabla^2 \right)
    \left( \Phi^{(1)} - \Psi^{(1)} \right),
\eea
the evolution equation (\ref{equ:E2}) can be written as follows,
\beq
\label{equ:hh}
h''_{ij} + 2 \cH h'_{ij} - \nabla^2 h_{ij}
= - 4 \hat{\cal T}_{ij}^{~lm} {\cal S}_{lm}\, ,
\eeq
where we have neglected the tensor part of $\Pi^{(2)i}_{~~~~j}$ and defined
\bea
{\cal S}_{ij}
&\equiv& 2 \Phi \partial^i \partial_j \Phi - 2 \Psi \partial^i \partial_j \Phi
    + 4 \Psi \partial^i \partial_j \Psi
    + \partial^i \Phi \partial_j \Phi - \partial^i \Phi \partial_j \Psi
    - \partial^i \Psi \partial_j \Phi + 3 \partial^i \Psi \partial_j \Psi
    \nonumber \\
& & - \frac{4}{3(1+w) \cH^2} \partial_i \left( \Psi' + \cH \Phi \right)
      \partial_j \left( \Psi' + \cH \Phi \right)
    - \frac{2 c_{s}^{2}}{3 w \cH^2} \left[ 3 \cH (\cH \Phi - \Psi') + \nabla^2 
\Psi \right]
      \partial_i \partial_j (\Phi - \Psi)\, .
\eea
Here, $w \equiv P^{(0)}/{\rho^{(0)}}$, $\Phi \equiv \Phi^{(1)}$ and $\Psi \equiv 
\Psi^{(1)}$.
We define the Fourier transform of tensor metric perturbations as 
\beq
h_{ij}(\mathbf{x},\eta) = \int \frac{\d^3 \mathbf{k}}{(2\pi)^{3/2}} e^{i 
\mathbf{k} \cdot \mathbf{x}} \left[ h_{\mathbf{k}}(\eta) 
\mathsf{e}_{ij}(\mathbf{k}) + \bar h_{\mathbf{k}}(\eta) \mathsf{\bar 
e}_{ij}(\mathbf{k}) \right]\, ,
\eeq
where the two time-independent polarization tensors $\mathsf{e}_{ij}$ and 
$\mathsf{\bar e}_{ij}$ may be expressed in terms of orthonormal basis vectors 
$\mathbf{e}$ and $ \bar{\mathbf{e}}$ orthogonal to $\mathbf{k}$,
\bea
\mathsf{e}_{ij}(\mathbf{k}) &\equiv& \frac{1}{\sqrt{2}} 
[\mathsf{e}_i(\mathbf{k}) \mathsf{e}_j(\mathbf{k}) - \mathsf{\bar 
e}_i(\mathbf{k}) \mathsf{\bar e}_j(\mathbf{k})]\, ,\\
\mathsf{\bar e}_{ij}(\mathbf{k}) &\equiv& \frac{1}{\sqrt{2}} 
[\mathsf{e}_i(\mathbf{k}) \mathsf{\bar e}_j(\mathbf{k}) + \mathsf{\bar 
e}_i(\mathbf{k}) \mathsf{e}_j(\mathbf{k})]\, .
\eea
In terms of these polarization tensors, the projection tensor in (\ref{equ:E2}) 
and (\ref{equ:hh}) is
\beq
 \hat{\cal T}_{ij}^{~lm} {\cal S}_{lm} = \int \frac{\d^3 
\mathbf{k}}{(2\pi)^{3/2}} e^{i \mathbf{k} \cdot \mathbf{x} } \left[ 
\mathsf{e}_{ij}({\mathbf k})   \mathsf{e}^{lm}(\mathbf{k}) + \mathsf{\bar 
e}_{ij}({\mathbf k})  \mathsf{ \bar e}^{lm}(\mathbf{k})\right] {\cal 
S}_{lm}(\mathbf{k})\, ,
\eeq
where
\beq
{\cal S}_{lm}(\mathbf{k}) =  \int \frac{\d^3 \mathbf{x'}}{(2\pi)^{3/2}} e^{- i 
\mathbf{k} \cdot \mathbf{x'} } {\cal S}_{lm}(\mathbf{x'})\, .
\eeq
In Fourier space, the equation of motion for the gravitational wave amplitude $h$
(for either polarization $h$ or $\bar h$) becomes
\beq
h''_{\mathbf{k}} + 2 \cH h'_{\mathbf{k}} + k^2 h_{\mathbf{k}}=  {\cal 
S}(\mathbf{k},\eta)\, ,
\label{EOMh}
\eeq
where the source term, ${\cal S}$, is a convolution of two first-order scalar perturbations at different wavenumbers,
\bea
\label{equ:source}
{\cal S}(\mathbf{k},\eta)
&=& -4 \mathsf{e}^{lm}(\mathbf{k} ) {\cal S}_{lm}(\mathbf{k})\\
&=& 4 \int \frac{\d^3 \mathbf{\tilde k}}{(2\pi)^{3/2}} 
\mathsf{e}^{lm}(\mathbf{k}) {\tilde k}_l {\tilde k}_m
    \Biggl[ \left\{ \frac{7+3w}{3(1+w)} - \frac{2 c_s^2}{w} \right\}
            \Phi_{\mathbf{\tilde k}}(\eta) \Phi_{\mathbf{k-\tilde k}}(\eta)
            + \left( 1 - \frac{2 c_s^2 {\tilde k}^2}{3 w \cH^2} \right)
              \Psi_{\mathbf{\tilde k}}(\eta) \Psi_{\mathbf{k-\tilde k}}(\eta) 
\nonumber \\
& &         + \frac{2 c_s^2}{w} \left( 1 + \frac{{\tilde k}^2}{3 \cH^2} \right)
              \Phi_{\mathbf{\tilde k}}(\eta) \Psi_{\mathbf{k-\tilde k}}(\eta)
            + \left\{ \frac{8}{3(1+w)} + \frac{2 c_s^2}{w} \right\} 
\frac{1}{\cH}
              \Phi_{\mathbf{\tilde k}}(\eta) \Psi'_{\mathbf{k-\tilde k}}(\eta) 
\nonumber \\
& &         - \frac{2 c_s^2}{w \cH}
              \Psi_{\mathbf{\tilde k}}(\eta) \Psi'_{\mathbf{k-\tilde k}}(\eta)
            + \frac{4}{3(1+w) \cH^2}
              \Psi'_{\mathbf{\tilde k}}(\eta) \Psi'_{\mathbf{k-\tilde k}}(\eta)
    \Biggr]\, . \label{equ:Source2}
\eea
Equation (\ref{equ:Source2}) reduces to the expression in \cite{Wands} in the limit $\Psi 
\rightarrow \Phi, w \rightarrow 1/3$
and $c_s^2 \rightarrow 1/3$.
The limit $\Psi \rightarrow \Phi, w \rightarrow 0$
and $c_s^2 \rightarrow 0$ was discussed in \cite{Mollerach}.

\subsection{Power Spectrum}

The power spectrum of tensor metric perturbations, $P_h(k,\eta)$, is defined as follows
\beq
\label{equ:Ph}
\langle h_{\mathbf{k}}(\eta) h_{\mathbf{K}}(\eta) \rangle = \frac{2 \pi^2}{k^3} 
\delta(\mathbf{k} + \mathbf{K}) P_h(k, \eta)\, .
\eeq
We now derive an expression for the power spectrum of second-order 
gravitational waves by solving equation (\ref{EOMh}). It is convenient to remove the Hubble damping term in (\ref{EOMh}) by defining $a h_\vk \equiv v_\vk$, where $v_\vk $ satisfies the following equation of motion
\beq
\label{equ:v}
v''_\vk + \Bigl( k^2 - \frac{a''}{a} \Bigr) v_\vk  = a {\cal S}\, .
\eeq
The particular solution of (\ref{EOMh}) is then found by the Green's 
function method
\beq
\label{equ:solh}
h_\vk(\eta) =  \frac{1}{a(\eta)} \int \d\tilde \eta \,  g_\vk(\eta;\tilde \eta) 
\Bigl[ a(\tilde \eta) {\cal S}(\vk,\tilde \eta) \Bigr]\, ,
\eeq
where 
\beq
\label{equ:g}
g''_\vk + \Bigl( k^2 - \frac{a''}{a} \Bigr) g_\vk  = \delta(\eta-\tilde\eta)\, .
\eeq
Exact solutions to (\ref{equ:g}) for both matter and radiation domination are 
derived in Appendix \ref{sec:Green}.
Substituting the solution (\ref{equ:solh}) into the expression for the tensor power spectrum (\ref{equ:Ph}) we find
\beq
\langle h_{\mathbf{k}}(\eta) h_{\mathbf{K}}(\eta) \rangle
= \frac{1}{a^2(\eta)} \int_{\eta_0}^\eta \d \tilde \eta_2 \int_{\eta_0}^\eta \d 
\tilde \eta_1 \,
  a(\tilde \eta_1) a(\tilde \eta_2) g_\vk(\eta; \tilde \eta_1) g_{\vK}(\eta; \tilde 
\eta_2)\,
  \langle {\cal S}(\mathbf{k}, \tilde \eta_1) {\cal S}(\mathbf{K}, \tilde 
\eta_2) \rangle\, .
\label{h-correlation}
\eeq
The source term (\ref{equ:Source2}) may be written in the following form
\beq
{\cal S}(\mathbf{k}, \eta) \equiv \int \d^3 \mathbf{\tilde k} \,  
\mathsf{e}(\mathbf{k}, \mathbf{\tilde k}) f(\mathbf{k}, \mathbf{\tilde{k}}, 
\eta) \, \psi_{\mathbf{k-\tilde k}} \psi_{\mathbf{\tilde k}}\, ,\eeq
where 
\beq
\mathsf{e}(\mathbf{k}, \mathbf{\tilde k}) \equiv \mathsf{e}^{ij}(k) 
\tilde k_i \tilde k_j =  \tilde k^2 [1-\mu^2]\, , \quad \mu \equiv \frac{\mathbf{k} 
\cdot \mathbf{\tilde k}}{k \tilde k}
\eeq
 and
\bea
\label{f}
f(\mathbf{k}, \mathbf{\tilde{k}}, \eta) &\equiv&
4 \Biggl[ \left\{ \frac{7+3w}{3(1+w)} - \frac{2 c_s^2}{w} \right\}
          \Phi({\tilde k}\eta) \Phi(|\mathbf{k-\tilde k}|\eta)
          + \left( 1 - \frac{2 c_s^2 {\tilde k}^2}{3 w \cH^2} \right)
            \Psi({\tilde k}\eta) \Psi(|\mathbf{k-\tilde k}|\eta) \nonumber \\
& &       + \frac{2 c_s^2}{w} \left( 1 + \frac{{\tilde k}^2}{3 \cH^2} \right)
            \Phi({\tilde k}\eta) \Psi(|\mathbf{k-\tilde k}|\eta)
          + \left\{ \frac{8}{3(1+w)} + \frac{2 c_s^2}{w} \right\} \frac{1}{\cH}
            \Phi({\tilde k}\eta) \Psi'(|\mathbf{k-\tilde k}|\eta) \nonumber \\
& &       - \frac{2 c_s^2}{w \cH}
            \Psi({\tilde k}\eta) \Psi'(|\mathbf{k-\tilde k}|\eta)
          + \frac{4}{3(1+w) \cH^2}
            \Psi'({\tilde k}\eta) \Psi'(|\mathbf{k-\tilde k}|\eta)
  \Biggr].
\eea
Here we have split the first-order quantities into transfer functions, $\Phi(k \eta)$, $\Psi(k \eta)$, and
primordial fluctuations $\psi_{\mathbf{k}}$,
\beq
\Phi_{\mathbf{k}}(\eta) \equiv \Phi(k\eta) \, \psi_{\mathbf{k}}, ~~~
\Psi_{\mathbf{k}}(\eta) \equiv \Psi(k\eta) \, \psi_{\mathbf{k}}\, .
\eeq
The primordial fluctuations are characterized by the power spectrum,
\beq
\langle \psi_{\mathbf{k}} \psi_{\mathbf{\tilde k}} \rangle
= \frac{2 \pi^2}{k^3} P(k)\, \delta(\mathbf{k} + \mathbf{\tilde k})\, .
\eeq
Observationally, it is found that $P(k)$ is nearly scale-invariant, so that the following parameterization is appropriate
\beq
P(k) = \frac{4}{9} \Delta_{\cal R}^2(k_0) \left( \frac{k}{k_0} \right)^{n_s -1}\, ,
\eeq
where recent CMB and large scale structure results \cite{WMAP, SDSS} imply $\Delta_{\cal R}^2(k_0=0.002\, {\rm Mpc}^{-1}) =(2.40 \pm 0.12) \times 10^{-9}$ and $n_s(k_0) \sim 0.94 - 1.10$ ($0.95 \pm 0.02$; no tensors).
Finally, the correlator in equation (\ref{h-correlation}) can be computed using 
Wick's theorem
\bea
\label{equ:SS}
\langle {\cal S}(\mathbf{k}, \tilde \eta_1) {\cal S}(\mathbf{K}, \tilde \eta_2) 
\rangle 
&=& \int \d^3 \mathbf{\tilde{k}} \, \mathsf{e}(\mathbf{k},\mathbf{\tilde k}) \, 
f(\mathbf{k},\mathbf{\tilde k},\tilde \eta_1)
    \int \d^3 \mathbf{\tilde{K}} \, \mathsf{e}(\mathbf{K},\mathbf{\tilde K}) \, 
f(\mathbf{K},\mathbf{\tilde K},\tilde \eta_2)\,
    \langle \psi_{\mathbf{k-\tilde k}} \psi_{\mathbf{\tilde k}} \psi_{\mathbf{K-
\tilde K}} \psi_{\mathbf{\tilde K}}
    \rangle 
\nonumber \\
&=& \delta(\mathbf{k}+\mathbf{K})  \int \d^3 \mathbf{\tilde{k}}\, 
\mathsf{e}(\mathbf{k}, \mathbf{\tilde k})^2  \,
    f(\mathbf{k}, \mathbf{\tilde k}, \tilde \eta_1)
    \Bigl[ f(\mathbf{k},\mathbf{\tilde k},\tilde \eta_2)+f(\mathbf{k},\mathbf{k-
\tilde k}, \tilde \eta_2) \Bigr]
    \frac{P(|\mathbf{k-\tilde k}|)}{|\mathbf{k - \tilde k}|^3} \frac{P(\tilde 
k)}{\tilde k^3}\, ,
\eea
and the power spectrum of second-order gravitational waves is
\beq
P_h(k, \eta)
= \int_0^\infty \d \tilde k  \int_{-1}^1 \d \mu \,
  P(|\mathbf{k-\tilde k}|) P(\tilde k) \, {\cal F}(k, \tilde k, \mu; \eta)\, ,
\eeq
where
\beq
{\cal F}(k, \tilde k, \mu; \eta) \equiv
\frac{[1-\mu^2]^2}{a^2(\eta)} \frac{k^3 \tilde k^3}{|\mathbf{k - \tilde k}|^3}
\int_{\eta_0}^\eta \d \tilde \eta_2 \d \tilde \eta_1 a(\tilde \eta_1) a(\tilde 
\eta_2)
g_k(\eta; \tilde \eta_1) g_k(\eta;\tilde \eta_2)
f(\mathbf{k}, \mathbf{\tilde k}, \tilde \eta_1)
\Bigl[ f(\mathbf{k},\mathbf{\tilde k}, \tilde \eta_2)+f(\mathbf{k},\mathbf{k-
\tilde k}, \tilde \eta_2) \Bigr] \, .
\eeq
Notice that the power spectrum $P_h(k, \eta)$ is defined completely in terms of the 
Green's function $g_\vk$ (Appendix \ref{sec:Green}),
the transfer functions $\Phi$ and $\Psi$ (Appendix \ref{sec:Bardeen}), and the 
primordial power spectrum of first-order scalar fluctuations, $P(k)$ (WMAP \cite{WMAP}).

\section{Analytical Description of the Spectrum}
\label{sec:analytic}

In this section we estimate the complete spectrum of scalar-induced gravitational 
waves analytically.
To simplify the analysis we neglect anisotropic stress and set $\Psi = \Phi$.
 In section \ref{sec:numeric} we evaluate the exact spectrum numerically 
including anisotropic stress and show that this gives only a small correction. With $\Psi = \Phi$,
 the source term of the equation of motion (\ref{EOMh}) can be expressed 
solely
by the Bardeen potential $\Phi$,
\beq
\label{equ:hhh}
h_k'' + 2 \cH h_k' + k^2 h_k=  {\cal S}(\Phi(k\eta))\, ,
\eeq
and $f(\mathbf{k}, \mathbf{\tilde{k}}, \eta)$ in equation (\ref{f}) is expressed
by a single transfer function $\Phi$,
\beq
\frac{3(1+w)}{4} f(\mathbf{k}, \mathbf{\tilde{k}}, \eta) =
2 (5+3 w) \Phi(|\mathbf{k-\tilde k}| \eta) \Phi(| \mathbf{\tilde k}| \eta)
+ 4 \left( 2 \eta \Phi(|\mathbf{k-\tilde k}|\eta) +  \eta^2 \Phi'(|\mathbf{k-
\tilde k}|\eta) \right)
  \Phi'(|\mathbf{\tilde k}|\eta)\, .
\eeq
In Appendix \ref{sec:Bardeen} we show that the transfer function for first-order 
scalar modes can be written in the following form
\beq
\label{equ:PhiX}
\Phi(k \eta) = \left\{ \begin{array}{cc} \frac{1}{1+ k^2 \eta^2} & \eta < 
\eta_{\rm eq}\\
\frac{1}{1+ k^2 \eta_{\rm eq}^2} & \eta > \eta_{\rm eq}
\end{array} \right. 
\eeq
\begin{figure}[h] 
   \centering
   \includegraphics[width=0.65\textwidth]{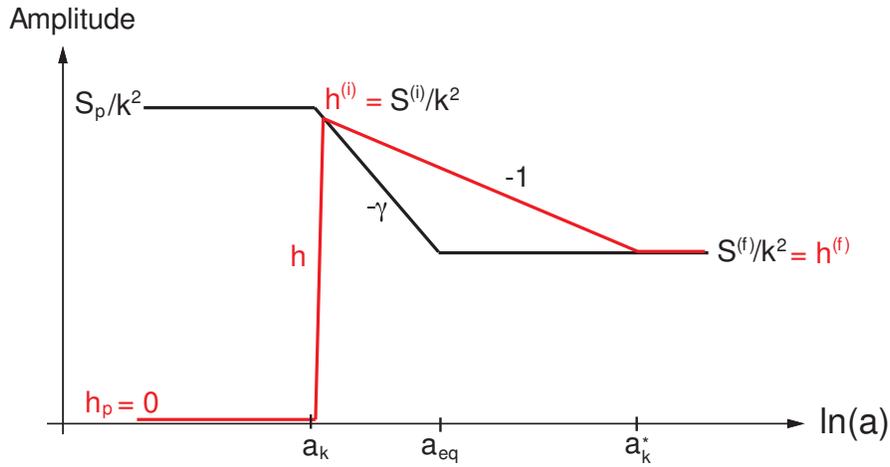} 
   \caption{{\bf Evolution of scalar source and induced gravitational waves.} 
Second-order tensors, $h$, are generated when the mode $k$ enters the horizon at 
$a_k$. If horizon entry occurs during the radiation dominated era, then the 
scalar source decays as $a^{-\gamma}$ until matter-radiation equality, $a_{\rm 
eq}$. During matter domination the scalar source terms remains at a constant 
value, ${\cal S}^{\rm (f)}$. Gravitational waves redshift like $a^{-1}$ as long 
as $h > {\cal S}^{\rm (f)}/k^2$, but remain at a constant amplitude maintained 
by the constant source term after that, $a > a^*_k$.}
   \label{fig:S}
\end{figure}

To study the generation of $h$ induced by ${\cal S}$ we make the approximation that
gravitational waves are produced instantaneously when the relevant mode enters the horizon.
The subsequent evolution of the tensor mode is scale-dependent and determined by 
the time evolution of the scalar source term (see Figure \ref{fig:S}). Scalar-induced gravitational waves 
redshift as long as
their magnitude is greater than ${\cal S}/k^2$. After that they freeze at a 
constant value maintained
by the constant source term during matter domination.
We define the transfer function for scalar-induced gravitational waves, $t(k, 
\eta)$, as follows
\beq
h_k(\eta) \equiv t(k,\eta) h_k^{\rm (i)}\, ,
\eeq
where $h_k^{\rm (i)}$ is the value of $h_k$ just after the instantaneous 
generation of gravitational waves after horizon entry (see Figure \ref{fig:S}). 
We estimate $h_k^{\rm (i)}$ by dropping time 
derivatives in the equation of motion (\ref{equ:hhh}) (since $k \eta > 1$ after horizon entry)
\beq
h_k^{\rm (i)} \sim \frac{1}{k^2} {\cal S}^{\rm (i)}\, .
\eeq
In \S\ref{sec:Phi} we calculate the initial power spectrum at the time of 
horizon crossing, 
\beq
P_h^{\rm (i)}(k,\eta_{\rm i}(k)) \equiv \frac{k^3}{2 \pi^2} \langle (h_k^{\rm 
(i)})^2\rangle \, ,
\eeq
where $\eta_{\rm i}(k) \sim k^{-1}$ is the conformal time when a comoving scale 
$k$ enters the horizon.
This initial spectrum is processed using the tensor transfer function, $t(k, 
\eta)$, which we derive in 
\S\ref{sec:source}.
Finally, in \S\ref{sec:spectrum}, we put these results together and compute the 
relative energy density of scalar-induced gravitational waves
\bea
\label{equ:Omega}
\Omega_{\rm GW}^{(2)}(k,\eta)
&=& \frac{1}{ 6 \pi^2 {\cal H}^2(\eta)} k^2 t^{2}(k,\eta) P^{\rm (i)}_{h}(k) 
\nonumber \\
&=& \frac{a(\eta) k^2}{a_{\rm eq} k_{\rm eq}^2} t^{2}(k,\eta) P^{\rm 
(i)}_{h}(k)\, .
\eea

\subsection{Power Spectrum at Horizon Crossing}
\label{sec:Phi}

In this section we estimate the $k$-scaling of the horizon power spectrum of 
scalar-induced gravitational waves. 
The  horizon 
amplitude $h_k^{\rm (i)}$ is estimated from the equation of motion 
(\ref{equ:hhh}) as follows
\beq
h_k^{\rm (i)} \sim \frac{1}{k^2} {\cal S}^{\rm (i)} \sim
\frac{1}{k^2} \int \d^3 \vtk \, \tk^2 (1 - \mu^2) \Phi(\tk \eta_{\rm i}) \Phi(|\vk-\vtk| 
\eta_{\rm i})  \psi_{\vtk} \psi_{\vk-\vtk} \, ,
\eeq
and its power spectrum is,
\bea
\langle h_k^{\rm (i)} h_K^{\rm (i)} \rangle
&\sim& \frac{1}{k^2 K^2} \int \d^3 \vtk  \d^3 \vtK \, \tk^2 \tK^2 (1 - \mu^2) (1 
- \tilde{\mu}^2)
       \Phi(\tk \eta_{\rm i}) \Phi(|\vk-\vtk| \eta_{\rm i}) \Phi(\tK \eta_{\rm i}) \Phi(|\vK-\vtK| \eta_{\rm i})
       \langle \psi_{\vtk} \psi_{\vk-\vtk} \psi_{\vtK} \psi_{\vK-\vtK} \rangle
\nonumber \\
&\sim& \frac{1}{k^4} \delta(\vk + \vK)
       \int \d^3 \vtk \, \tk^4 (1 - \mu^2)^2 \Phi^2(\tk \eta_{\rm i}) \Phi^2(|\vk-\vtk| 
\eta_{\rm i})
            \frac{P(\tk)}{\tk^3} \frac{P(|\vk-\vtk|)}{|\vk-\vtk|^3}\, .
\eea
Hence,
\beq
P_h^{\rm (i)}(k) \equiv \frac{k^3}{2 \pi^2} \langle (h_k^{({\rm i})})^2 \rangle 
\sim
\frac{1}{k} \int \d^3 \vtk \, \tk^4 (1 - \mu^2)^2 \Phi^2(\tk \eta_{\rm i}) \Phi^2(|\vk-
\vtk| \eta_{\rm i})
\frac{P(\tk)}{\tk^3} \frac{P(|\vk-\vtk|)}{|\vk-\vtk|^3}\, . \label{equ:phi0}
\eeq
To compute (\ref{equ:phi0}) we use the transfer function for the scalar 
potential (\ref{equ:PhiX})
and assume a scale-invariant spectrum, $P(k) = \frac{4}{9} \Delta_{\cal R}^2$. (The scale-dependence of the scalar spectrum can be reinserted at the end of the computation).
Here we have defined $\Delta_{\cal R}^2 \approx 10^{-9}$ as a measure of scalar 
power on COBE scales $k_0 \approx 0.002\, {\rm Mpc}^{-1}$.
Hence,
for the radiation dominated phase, we have,
\bea
P^{\rm (i)}_{h}(k,\eta_{\rm i}(k))
&\sim&
\frac{\Delta_{\cal R}^4}{k} \int^{\infty}_{0} \d \tk \int^{1}_{-1} \d \mu (1 - 
\mu^2)^2
\frac{\tk^3}{(k^2 + \tk^2 - 2 k \tk \mu)^{3/2}} \frac{1}{[1 + (\tk/k)^2]^2}
\frac{1}{[1 + (k^2 + \tk^2 - 2 k \tk \mu)/k^2]^2} \nonumber \\
&=&
\Delta_{\cal R}^4 \int^{\infty}_{0} \d x \int^{1}_{-1} \d\mu (1 - \mu^2)^2
\frac{x^3}{(1 + x^2 - 2 x \mu)^{3/2}} \frac{1}{(1 + x^2)^2} \frac{1}{(2 + x^2 - 
2 x \mu)^2} \nonumber \\
&\propto& \Delta_{\cal R}^4\, ,
\eea
where we defined $x \equiv \tk /k$. On the other hand, for the matter dominated 
phase, we have,
\bea
P^{\rm (i)}_{h}(k,\eta_{\rm i}(k))
&\sim&
\frac{\Delta_{\cal R}^4}{k} \int^{\infty}_{0} \d \tk \int^{1}_{-1} \d\mu (1 - 
\mu^2)^2
\frac{\tk^3}{(k^2 + \tk^2 - 2 k \tk \mu)^{3/2}} \frac{1}{[1 + (\tk/k_{\rm 
eq})^2]^2}
\frac{1}{[1 + (k^2 + \tk^2 - 2 k \tk \mu)/k_{\rm eq}^2]^2} \nonumber \\
&=&
\Delta_{\cal R}^4 \int^{\infty}_{0} \d x \int^{1}_{-1} \d\mu (1 - \mu^2)^2
\frac{x^3}{(1 + x^2 - 2 x \mu)^{3/2}} \frac{1}{(1 + x^2 y^2)^2} \frac{1}{[1+ (1 
+ x^2 - 2 x \mu) y^2]^2} \, ,
\eea
where $x \equiv \tk /k$ and $y \equiv k/k_{\rm eq} \ll 1$. Neglecting the $\mu$-terms we find
\bea
P^{\rm (i)}_{h}(k,\eta_{\rm i}(k)) &\sim&
\Delta_{\cal R}^4 \int^{\infty}_{0} \d x
\frac{x^3}{(1 + x^2)^{3/2}} \frac{1}{(1 + x^2 y^2)^4} \nonumber \\
&=&
\Delta_{\cal R}^4
\left[ \frac{5 (1 + 6 y^2) \arccos{y}}{16 y (1 - y^2)^{9/2}}
       - \frac{81 + 28 y^2 - 4 y^4}{48 (1 - y^2)^4}
\right] \nonumber \\
&\sim& \frac{5 \pi \Delta_{\cal R}^4}{32 y} \nonumber \\
&\propto& \Delta_{\cal R}^4 \frac{k_{\rm eq}}{k}\, .
\eea
The power spectrum at horizon crossing therefore scales as follows
\beq
\label{equ:FinalPh}
P^{\rm (i)}_h(k) \propto \Delta_{\cal R}^4 \left\{ \begin{array}{l l} 
\frac{k_{\rm eq}}{k} & \ \quad k < k_{\rm eq} \\  1 & \ \quad  k > k_{\rm eq} 
\end{array} \right.
\eeq

\subsection{Second-Order Tensor Transfer Function}
\label{sec:source}

To compute the transfer function for second-order, scalar-induced gravitational waves we need to 
estimate the time evolution of the source term.
In particular, the transfer function for modes that enter the horizon during radiation domination 
is sensitive only to ratio of the source terms at horizon crossing, ${\cal 
S}^{\rm (i)}$ and the asymptotic value after equality, ${\cal S}^{\rm (f)}$ (see 
Figure \ref{fig:S}).
We parameterize the decay of the source term during radiation domination as follows
\beq
\label{equ:ratio}
\frac{{\cal S}^{\rm (f)}}{{\cal S}^{\rm (i)}} = \left( \frac{a_k}{a_{\rm 
eq}}\right)^{\gamma(k)}\, ,
\eeq
where we have allowed for a scale-dependence of the effective decay rate.
In the following we put limits on $\gamma(k)$ by considering the asymptotic 
evolution of subhorizon modes ($k \eta \gg 1$).\\

Let $x \equiv |\mathbf{k -\tilde k}| \eta$ and $y \equiv |\mathbf{\tilde k}| 
\eta$.
Using the following relations
\beq
\Bigl( \frac{x(y,\mu)}{k \eta} \Bigr)^2
= 1 + \Bigl( \frac{y}{k \eta} \Bigr)^2 - 2\Bigl( \frac{y}{k \eta} \Bigr) \mu\,
\quad \quad  {\rm and} \quad \quad
\psi_{\mathbf{k-\tilde k}} \psi_{\mathbf{\tilde k}} \propto 
\frac{\eta^3}{x^{3/2} y^{3/2}}\, ,
\eeq
the source term may be written as follows
\beq
\label{equ:int}
{\cal S} (\mathbf{k}, \eta)
\propto \frac{2 \pi}{\eta^2} \int_0^\infty \d y \int_{-1}^1 \d \mu\, [1-\mu^2] 
\,
        \frac{y^{5/2}}{x(y,\mu)^{3/2}} \, f(x,y)\, ,
\eeq
where
\bea
2 f(x,y) &=& \Phi(x) \Phi(y) \Bigl[ 6 - y^2 \Phi(x) \Phi(y)\Bigr]\, , \\
&=& \frac{1}{x^2+1} \frac{1}{y^2+1} \left[ 6 - \frac{1}{x^2+1} 
\frac{y^2}{y^2+1}\right] \, .
\eea
Let us estimate the integral (\ref{equ:int}). 
The limit $y \to 0$ is clearly suppressed by the phase space factor $y^{5/2}$ in 
the integrand.
The limit $x \to 0$ ($y \to k\eta$, $\mu \to 1$) is suppressed by the projection 
factor $[1-\mu^2]$.
To see this, first take the limit $y \to k \eta$,
\beq
\Bigl(\frac{x}{k \eta} \Bigr)^2 \to 2 [1-\mu]\, , \quad
\frac{1-\mu^2}{x^{3/2}} \to (1-\mu)^{1/4} (1+\mu)\, .
\eeq
This shows that the integrand vanishes in the limit $x \to 0$. In addition, 
large $x$ and $y$ are suppressed by the transfer function $f(x,y)$ ({\it i.e.} 
the decay of the Bardeen potential on subhorizon scales).
The dominant contribution to the integral (\ref{equ:int}) therefore comes from 
regions of phase space where $\tilde k \sim k$ ($y \sim k \eta$) and 
$|\mathbf{k-\tilde k}| \sim k$ ($x \sim k \eta$, $\mu \sim 0$).
Let us therefore write
\beq
{\cal S} \propto \frac{1}{\eta^2} \int \d \ln y \int \d \ln (1-\mu) \, (1-\mu)^2 
(1+\mu) \frac{y^{7/2}}{x^{3/2}} \frac{1}{x^2+1} \frac{1}{y^2+1} \left[ 6 - 
\frac{1}{x^2+1} \frac{y^2}{y^2+1}\right]
\eeq
and take the subhorizon limit $y,x \to k \eta > 1$, $\mu \to 0$
\beq
{\cal S} \propto \frac{1}{\eta^2} \frac{(k\eta)^{7/2}}{(k \eta)^{3/2}} 
\frac{1}{[(k \eta)^2 +1]^2} \left[6 -  \frac{(k \eta)^2}{[(k \eta)^2 +1]^2} 
\right] \approx \frac{1}{\eta^2} \frac{1}{(k \eta)^2} \propto \frac{1}{a^4}\, .
\eeq
Hence the source term decays at most as $a^{-4}$ after the mode $k$ enters the 
horizon during the radiation dominated era, {\it i.e.} $\gamma(k) < 4$.
In fact, we expect the source term to decay considerably slower than that for a 
while after horizon crossing.
The source will decay more and more quickly as 
the horizon grows much larger than the wavelength of the mode,
finally reaching the asymptotic behavior that is proportional to $a^{-4}$. This 
leads us to expect that the effective $\gamma$ in equation (\ref{equ:ratio}) 
will be significantly smaller than 4.
Numerically, we find $\gamma \approx 3$ (see \S\ref{sec:numeric}).\\

The transfer function for second-order gravitational waves is considerably different from 
the transfer function for first-order gravitational waves.
First of all, modes which enter the horizon during matter domination have constant source terms 
and hence do {\it not} decay
\beq
t(k, \eta) = 1\, , \hspace{1cm} k < k_{\rm eq}\, .
\eeq
Next, consider the evolution of the scalar source term and induced gravitational waves for modes 
that enter the horizon during the radiation dominated era (see Figure \ref{fig:S}).
Here,
$k=a_k H$ defines the time of horizon entry ($a_k$) for a mode of wavenumber 
$k$. 
We assume that $h$ grows very rapidly after horizon entry to become of order the 
source term.
Then ${\cal S}$ decays as $a^{-\gamma}$ (where our previous discussion implies 
$\gamma < 4$) while $h$ redshifts as $a^{-1}$ until $h$ is equal to the final 
source term during matter domination at $a^*_k$. For $a> a^*_k$, $h$ stays constant.
We therefore have,
\beq
\frac{h^{\rm (f)}}{h^{\rm (i)}} = \frac{a_k}{a^*_k} \approx \frac{{\cal S}^{\rm 
(f)}}{{\cal S}^{\rm (i)}} = \Bigl( \frac{a_k}{a_{\rm eq}} \Bigr)^\gamma \, ,
\eeq
and find
\beq
\label{equ:xxx}
\frac{a_k}{a^*_k} = \Bigl( \frac{a_k}{a_{\rm eq}}\Bigr)^{\gamma} =  \Bigl( 
\frac{k}{k_{\rm eq}}\Bigr)^{-\gamma}\, .
\eeq
For a fixed time $\eta$, subhorizon modes with sufficiently large $k$ have never 
settled down. The critical wave number at a time $\eta$ can be obtained by 
substituting $a_k^* = a(\eta)$ into equation (\ref{equ:xxx}),
\beq
k_c(\eta) = \Bigl( \frac{a(\eta)}{a_{\rm eq}}\Bigr)^{1/(\gamma-1)} k_{\rm eq}\, 
.
\eeq
Modes with $k > k_c(\eta)$ simply redshift like $a^{-1}$,
\beq
t(k, \eta) = \frac{a_k}{a(\eta)} = \frac{a_{\rm eq}}{a(\eta)} \frac{1}{k 
\eta_{\rm eq}}\, , \hspace{1cm} k > k_c(\eta)\, .
\eeq
The transfer function for second-order gravitational waves therefore takes the 
following interesting form
\beq
\label{equ:Finalt}
t(k, \eta) = \left\{ \begin{array}{l l} 1 & \ \quad k < k_{\rm eq} \\  \Bigl( 
\frac{k}{k_{\rm eq}}\Bigr)^{-\gamma} & \ \quad k_{\rm eq} < k < k_c(\eta) \\  
\frac{a_{\rm eq}}{a(\eta)} \frac{k_{\rm eq}}{k} & \ \quad  k > k_c(\eta) 
\end{array} \right.
\eeq

\subsection{Spectrum of Scalar-Induced Gravitational Waves}
\label{sec:spectrum}

Substituting the power spectrum at horizon crossing (\ref{equ:FinalPh}) and the 
tensor transfer function (\ref{equ:Finalt}) into equation (\ref{equ:Omega}) for 
the relative spectral energy density of gravitational waves at time $\eta$, we 
find 
\beq
\Omega_{\rm GW}^{\rm (2)}(k, \eta)  = A_{\rm GW}^{(2)} \Delta_{\cal R}^4(k_0) \Bigl( \frac{k}{k_0} 
\Bigr)^{2(n_s -1)} \, \left\{ \begin{array}{l l} 
\frac{a(\eta)}{a_{\rm eq}}  \frac{k}{k_{\rm eq}}   & \ \quad k < k_{\rm eq} \\  
\frac{a(\eta)}{a_{\rm eq}} \Bigl( \frac{k}{k_{\rm eq}}\Bigr)^{-(2\gamma-2)} & \ 
\quad k_{\rm eq} < k < k_c(\eta) \\  \frac{a_{\rm eq}}{a(\eta)} & \ \quad  k > 
k_c(\eta) 
\end{array} \right.
\eeq
where an overall normalization constant $A_{\rm GW}^{(2)}$ hasn't been fixed by our analytical arguments. In \S\ref{sec:numeric} we find $A_{\rm GW}^{(2)} \approx 10$ and $\gamma \approx 3$ (This is consistent with the normalization of the small-scale spectrum in \cite{Wands}).
Figure \ref{fig:1} summarizes this conjectured form of the scalar-induced gravitational wave spectrum. 
If the scalar spectrum can be treated by a power law with constant spectral index over a large range of scales, then a blue scalar spectrum ($n_s > 1$) enhances the tensor spectrum on small scales, while a red spectrum ($n_s < 1$) suppresses secondary tensor fluctuations. 
For comparison, the first-order spectrum of primordial gravitational waves in inflationary models can be expressed as 
\beq
\Omega_{\rm GW}^{\rm (1)}(k, \eta)   = A_{\rm GW}^{(1)}  r_0  
\Delta_{\cal R}^2(k_0) 
\Bigl(\frac{k}{k_0}\Bigr)^{n_t} 
\, \left\{ 
\begin{array}{l l} 
 \frac{a_{\rm eq}}{a(\eta)}  \Bigl( \frac{k}{k_{\rm eq}} \Bigr)^{-2}  & \ \quad 
k < k_{\rm eq} \\  
 \frac{a_{\rm eq}}{a(\eta)} & \ \quad  k > k_{\rm eq} \end{array} \right.
\eeq
where $A_{\rm GW}^{(1)} = 4.2 \times 10^{-2}$ and $r_0 \equiv \frac{P_h(k_0)}{P(k_0)}$ is the first-order tensor-to-scalar ratio evaluated on the scale of today's horizon, $k=k_0 \approx 0.002\, {\rm Mpc}^{-1}$.
For single-field inflation, the spectral index of the primordial tensor spectrum, $n_t$, is related to the tensor-to-scalar ratio by the slow-roll consistency relation, $n_t = - r_0/8$. 
Matter-radiation equality is normalized by recent observations  \cite{WMAP} $ a_{\rm eq} \approx a_0/3400$.\\

The current tensor spectrum ($\eta = \eta_0$) on large scales 
($k=k_{\rm eq}$) and on very small scales ($k \gg k_{\rm eq}$) (assuming inflation and $n_t \approx 
0$ and $n_s \approx 1$) satisfies
\beq
\frac{\Omega_{\rm GW}^{(1)}(k=k_{\rm eq})}{\Omega_{\rm GW}^{(2)}(k=k_{\rm eq})} 
\approx r_0 \frac{A_{\rm GW}^{(1)}}{A_{\rm GW}^{(2)}}  \frac{\Bigl( \frac{a_{\rm eq}}{a_0}\Bigr)^2}{\Delta_{\cal R}^2(k_0)} \sim 
\frac{1}{10} r_0
\eeq
and
\beq
\frac{\Omega_{\rm GW}^{(1)}(k\gg k_{\rm eq})}{\Omega_{\rm GW}^{(2)}(k \gg k_{\rm 
eq})} \approx r_0 \frac{A_{\rm GW}^{(1)}}{A_{\rm GW}^{(2)}}  \frac{1}{\Delta_{\cal R}^2(k_0)} \sim 10^6\, r_0\, .
\eeq
Hence, for inflation, on large scales the second-order, scalar-induced contribution {\it today}, in fact, 
dominates over the first-order contribution. This reflects the 
fact that second-order gravitational waves do {\it not} redshift on large 
scales, while first-order gravitational waves redshift on all scales.
On small scales the first-order contribution dominates unless $r_0 < 10^{-6}$.
For ekpyrotic/cyclic models, the first-order contribution (due to direct quantum fluctuations of the metric) is suppressed at $k=k_0$ by 60 orders of magnitude compared to the inflationary signal \cite{LathamCyclic} and the spectrum is blue. Hence, in these models, the scalar-induced tensor modes, $\Omega_{\rm GW}^{(2)}$, comprise the dominant contribution on all scales.\\

CMB observations probe the time of photon-baryon decoupling at $a_{\rm CMB} 
\approx 3\, a_{\rm eq}$ and scales with $k < k_{\rm CMB} = \left( \frac{a_{\rm eq}}{a_{\rm CMB}} \right)^{1/2} k_{\rm eq}$.
The gravitational wave spectrum at that time satisfies 
\beq
\frac{\Omega_{\rm GW}^{(1)}(k=k_{\rm CMB})}{\Omega_{\rm GW}^{(2)}(k=k_{\rm CMB})} 
\approx r_0 \frac{A_{\rm GW}^{(1)}}{A_{\rm GW}^{(2)}} \frac{\Bigl( \frac{a_{\rm eq}}{a_{\rm CMB}}\Bigr)^2}{\Delta_{\cal 
R}^2(k_0)} \left( \frac{k_{\rm eq}}{k_{\rm CMB}} \right)^{3} \sim 10^6\, r_0
\eeq
and
\beq
\frac{\Omega_{\rm GW}^{(1)}(k\gg k_{\rm eq})}{\Omega_{\rm GW}^{(2)}(k \gg k_{\rm 
eq})} \approx r_0 \frac{A_{\rm GW}^{(1)}}{A_{\rm GW}^{(2)}} \frac{1}{\Delta_{\cal R}^2(k_0)} \sim 10^6\, r_0\, .
\eeq
Hence, at decoupling the first-order tensor signal dominates over the second-order signal if $r_0 > 10^{-6}$. (This is consistent with the result of 
Mollerach {\it et al.} \cite{Mollerach} who claim that second-order 
gravitational waves only have a significant imprint on the CMB if $r_0 < 10^{-
8}$. This corresponds to second-order gravitational waves dominating over first-order gravitational waves at the time of recombination. Second-order effects can become 
visible for larger $r_0 < 10^{-6}$ if late time polarization generated by 
reionization is considered \cite{Mollerach}.) \\

For completeness, let us consider the power spectrum on superhorizon 
scales, {\it e.g.} during the matter dominated phase.
On superhorizon scales, $k \ll {\cal H}$, the initial amplitude can be
estimated from the equation of motion (\ref{equ:hhh}) by ignoring the gradient 
term and approximating the time derivatives by factor of ${\cal H}$,
\beq
{\cal H}^2 h_{k \ll {\cal H}}^{\rm (i)} \sim {\cal S}\, ,
\eeq
while  $k^2 h_k^{\rm (i)} \sim {\cal S}$ estimates
the initial amplitude on the horizon scale. Hence, the initial power spectrum, 
$P_h^{\rm (i)}$, on superhorizon scales
is simply $(k/{\cal H})^4$ times the spectrum on the horizon scale
\beq
P_h^{\rm (i)}(k) \propto \left(\frac{k}{\cal H} \right)^4 \times \Delta_{\cal R}^4(k_0) 
\frac{k_{\rm eq}}{k} \propto k^3 \, , \quad \quad k < k_{\rm hor} \le k_{\rm 
eq}\, .
\eeq
Although the tensor power spectrum, $P_h$, is a well-defined gauge-invariant object on super-horizon scales, $\Omega_{\rm GW}^{(2)}$ is {\it not}. 
In particular, equation (\ref{equ:Omega}) is only defined on subhorizon scales.
Nevertheless, we formally {\it define} $\Omega_{\rm GW}^{(2)} \propto k^2 P_h^{\rm (i)} \propto k^5$ on superhorizon scales, but do {\it not} attribute physical meaning to it. This definition is useful, since all our results are presented in terms of $\Omega_{\rm GW}^{(2)}$ and the shape of the superhorizon spectrum gives a simple consistency check for the numerical analysis.

\section{Numerical Results for the Exact Spectrum}
\label{sec:numeric}

The spectrum of scalar-induced gravitational waves that we derived in \S\ref{sec:2t} and discussed analytically in \S\ref{sec:analytic} can be evaluated exactly using standard numerical methods.  The time evolution of the first-order perturbation variables necessary 
to compute the spectrum, $\Phi(k\eta)$ and $\Psi(k \eta)$, 
is obtained from publicly available Einstein-Boltzmann codes
such as CMBFAST \cite{CMBFAST} or CAMB \cite{CAMB}. 
We first store the time evolution of $\Phi$ and $\Psi$ in  $k$-space, then convolve them according to 
equation (\ref{equ:SS}).  In practice, the range of $k$ is taken be $[10^{-5}\, {\rm Mpc}^{-1}, 500\, {\rm Mpc}^{-1}]$ and variables are evaluated at 50 uniformly spaced points per log-interval of $k$.
We have checked that our results are stable under variations of the $k$-space boundaries 
and the discretization.

\begin{figure}[h] 
   \centering
   \includegraphics[width=0.75\textwidth]{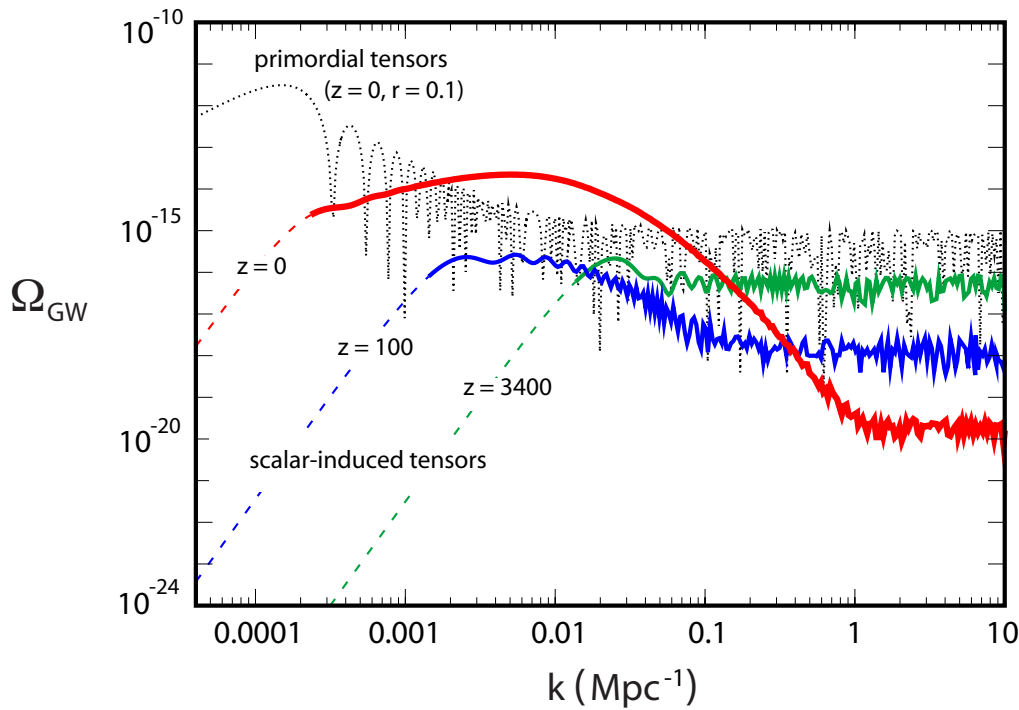} 
   \caption{Numerical spectra of scalar-induced gravitational waves (lower curves) and the scale-invariant primordial tensor spectrum for an inflationary model with tensor-to-scalar ratio $r=0.1$ (upper curve). The scalar-induced spectra are shown at three different epochs, $z+1= 3400, 100,$ and $1$.  Each curve has been extended, for pedagogical reasons, to modes with small wavenumbers $k$ that lie outside the horizon at the given epoch (dotted range of the three lower curves).  Note that  current ($z+1=1$) scalar-induced contributions cross the primordial inflationary contribution at intermediate wavelengths, as suggested by the schematic in Figure \ref{fig:1}. 
   The simulation assumes a flat $\Lambda$CDM cosmology with the following model parameters: $\Delta_{\cal R}^2(k_0=0.002\, {\rm Mpc}^{-1}) = 2.4 \times 10^{-9}$, $n_s = 1$, $n_t=0$, $r=0.1$, $\Omega_b h^2 = 0.022$, $\Omega_m h^2 =0.11$, $h=0.7$.  }
     \label{fig:num3}
\end{figure}

In the numerical analysis it is possible to incorporate 
the difference between $\Phi$ and $\Psi$ resulting from anisotropic 
stress of the fluid because the Boltzmann equations of photons and neutrinos
are solved explicitly in the code by expanding their distribution functions 
into multipole moments. (Neglecting anisotropic stress from neutrinos implies
$\sim$10\% errors for both first-order scalar and tensor perturbations 
\cite{Hu:1995fq,Weinberg:2003ur}; see Figure \ref{fig:trans} in Appendix \ref{sec:Bardeen}. For second-order tensors we find that the inclusion of anisotropic stress typically has less than 1\% effect on the amplitude of the spectrum.)
Finally, we should mention here that our definition of $c_s^2$, equation (\ref{equ:cs}), relates first-order pressure and energy density perturbations in the total matter, $P^{(1)}$ and 
$\rho^{(1)}$, including entropy perturbations, 
and thus it is scale-dependent. We have derived and incorporated this 
numerically.

The numerically calculated spectra are shown in Figure \ref{fig:num3}. 
The shape of the scalar-induced gravitational wave spectrum agrees well with the analytical results of the previous section and the schematic diagram in Figure \ref{fig:1}, except for the fine-scale oscillations and the modest smoothing of the enhanced feature at large wavelengths.  The scalar-induced spectrum is derived directly from observations of the scalar perturbations plus general relativity and is, in this sense, independent of the cosmological model for generating the primordial perturbations, {\it e.g.}, inflation vs. ekpyrotic/cyclic.  However, the transfer function does depend weakly on the expansion rate and composition of the universe, and, hence, the cosmological background parameters must be measured or otherwise specified.

\section{Discussion}
\label{sec:discussion}

Precise cosmological observations \cite{WMAP, SDSS} have confirmed the existence of a nearly scale-invariant spectrum of primordial scalar fluctuations.
These scalar fluctuations induce a second-order contribution to the 
spectrum of tensor perturbations that must be present for any cosmological model 
that accounts for the observed scalar spectrum. 
In particular, the computation of the second-order gravitational wave 
signal does {\it not} assume that the primordial perturbation spectra were 
generated by inflation -- it only relies on the observed spectrum of scalar perturbations and general relativity.

In this paper we have computed the scalar-induced spectrum of gravitational waves 
produced in the early Universe and used the Einstein equations to evolve it to 
the present.
We have extended previous approaches to this problem \cite{Mollerach, Wands} by considering the complete cosmic history for the evolution of 
second-order gravitational waves and allowing for anisotropic stress in our 
numerical work.

\begin{figure}[h] 
   \centering
   \includegraphics[width=0.63\textwidth]{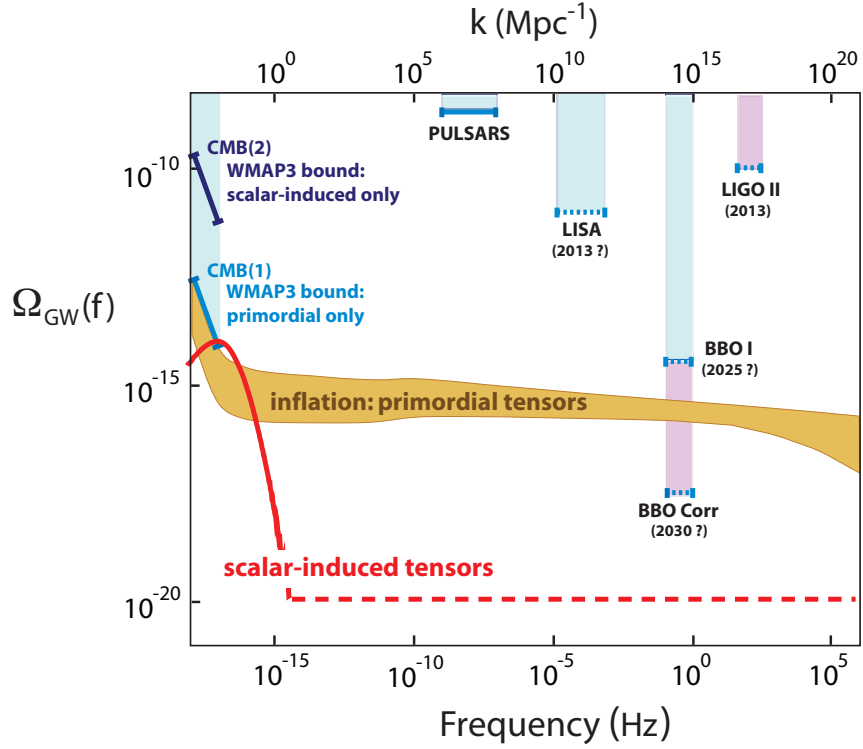} 
   \caption{{\bf Observational prospects.} Shown are the theoretical predictions for the present spectrum of primordial inflationary gravitational waves and scalar-induced gravitational waves, as well as current (solid bars) and future (dashed bars) experimental bounds (figure modified from \cite{Latham}). 
The range of amplitudes for the primordial tensors corresponds to minimally tuned models of inflation as described in \cite{Latham}. With more fine-tuning this signal can be suppressed. In contrast, the amplitude of the scalar-induced tensors is fixed by the observed amplitude of scalar fluctuations and therefore provides an absolute lower limit on the stochastic gravitational wave background.    
The CMB constraints depend on assumptions about the transfer function for gravitational waves to extrapolate constraints obtained at decoupling to the current spectrum. Since second-order gravitational waves do not redshift on CMB scales, the CMB observations imply separate constraints on the current first- and second-order spectra. These constraints are labeled CMB(1) and CMB(2), respectively.
The dashed section of the scalar-induced tensor spectrum illustrates extrapolation from CMB to direct-detection scales using a scale-invariant scalar spectrum ($n_s=1$). Important uncertainties in the extrapolation between CMB and BBO scales are discussed in the main text.}
   \label{fig:latham}
\end{figure}

Perhaps the most interesting theoretical feature is that second-order gravitational waves do {\it not} 
redshift on large and intermediate scales, but are maintained at a constant amplitude by the scalar source terms. 
This leads to a transfer function for the scalar-induced gravitational waves 
that produces a (nearly) scale-invariant spectrum on small scales and interesting scale-dependence on large and intermediate scales.
In particular, there is a peak in the current spectrum of second-order 
gravitational waves at the scale of the comoving horizon at matter-radiation 
equality (see Figures 
\ref{fig:1}, \ref{fig:num3} and \ref{fig:latham}) that is likely to exceed the primordial tensor spectrum at the present epoch.  Unfortunately, there are no known methods for directly probing the present gravitational wave spectrum
on these scales (corresponding to the size of superclusters today). At earlier times, such as recombination, the feature was much smaller and so only has small effects, {\it e.g.} on the CMB \cite{Mollerach}. Hence, this substantial feature is likely to remain of purely academic interest in the foreseeable future. 

On much smaller scales, which may be accessible to space-based laser-interferometer experiments (for a nice discussion see Ref.~\cite{LathamThesis}), there are no measurements of the scalar perturbation spectrum, so one must rely on extrapolating from what is known about scalar perturbations on large scales, {\it e.g.} from measurements of the CMB and large scale structure. Since these two wavelength regimes are separated by 16 orders of magnitude, extrapolation uncertainties can have important effects.
For example, the dashed line in Figure \ref{fig:latham}, which illustrates the extrapolation of the tensor spectrum based on a perfectly scale-invariant ($n_s=1$) scalar spectrum, is an estimate of the induced small-scale tensor signal.
Assuming that a nearly constant spectral index is a valid approximation from CMB scales to the smallest scales, a blue spectrum ($n_s > 1$) enhances scalar-induced tensor modes on direct-detection scales, while a red spectrum ($n_s < 1$) suppresses them.  
One might imagine improving on this simplistic extrapolation by including running of the scalar spectral index, $\alpha_s \equiv d n_s/ d \ln k$, although observationally $\alpha_s$ is only weakly constrained. 
However, there are interesting cases where this approach is inadequate \cite{LathamThesis} and can even lead to qualitatively misleading conclusions \cite{DBKT}.
Alternatively, for explicit inflationary models the whole spectrum can be computed directly from the inflaton potential $V(\phi)$ without expanding with respect to the CMB scale.\footnote{For the first-order inflationary tensor spectrum this approach was followed by \cite{Smith}.} 
In particular, for hybrid inflation models of the form $V(\phi) = V_0 \left[1 +  \left( \frac{\phi}{\mu} \right)^p \right]$, where $\mu > M_P$ and $\phi < \mu$, the exact computation predicts enhancement of the scalar-induced tensors on direct-detection scales, while the power law approximation (for $p \ge 3$) with $n_s > 1$ and $\alpha_s < 0$ predicts an overall suppression \cite{DBKT}.
In a follow-up paper, we explore examples of this type that may produce large signals at BBO scales \cite{followup}.\\

Finally, let us consider the possible scenarios for future observations of the 
first- and second-order tensor signals and what they would signify: 
\begin{itemize}
\item If a spectrum of gravitational waves is observed that conforms to a 
nearly scale-invariant, first-order tensor signal and with $r> 10^{-2}$ (as shown 
in Figure \ref{fig:latham}), this would be a spectacular confirmation of the inflationary model 
of the universe and completely rule out ekpyrotic/cyclic models. The observation 
of such a signal with $r < 10^{-2}$ would also rule out ekpyrotic/cyclic models 
but, in addition, inflationary models would be pushed into a fine-tuned 
region of parameter space \cite{Latham}. 
\item If no nearly scale-invariant, first-order tensor spectrum is detected
but the scalar-induced, second-order tensor spectrum is observed (either by 
extremely sensitive CMB polarization experiments or small scale direct detection 
interferometers like BBO), then inflation could only be compatible with extreme 
enough fine-tuning to suppress the first-order contribution to the tensor 
signal, and alternatives like the ekpyrotic/cyclic models would be favored.  
\item If future experiments show that there is no tensor signal at or above the 
level of the predicted scalar-induced tensor spectrum, either general relativity 
or the interpretation of the scalar fluctuations would have to be amiss.
\end{itemize} 
In practice, observing the second-order tensor signal is a long way off, at 
best, and perhaps even impossible given our current understanding of 
astrophysical foregrounds and detector limitations for both CMB and direct 
detection experiments \cite{CMB, DavidLyman}.
Nevertheless, we consider it interesting that the {\it observed} level of scalar 
fluctuations implies a 
model-independent {\it lower limit} on gravitational waves from the early 
Universe whose detailed features can be computed from a general relativistic 
description of cosmic evolution.

\acknowledgements
It is a pleasure to thank Latham Boyle for a very careful reading of a draft of this paper.
PJS and DB thank Jo Dunkley and David Spergel for initial discussions that helped inspire this work.
KI thanks Kouji Nakamura for useful discussions on second-order 
perturbations. KI and KT acknowledge the support by a Grant-in-Aid for the 
Japan Society for the Promotion of Science.
 PJS and DB are supported by US Department of Energy grant DE-FG02-91ER40671.
 
\newpage 
\appendix
\section{Green's Function for Gravitational Waves}
\label{sec:Green}

In this section we derive exact Green's functions of equation (\ref{EOMh}) for 
both the radiation and matter dominated eras.
The Green's function of a general second-order differential equation, $\hat {\cal 
L} g = \delta(\eta - \tilde \eta)$ is defined as follows
\beq
g(\eta;\tilde \eta) \equiv \frac{v_1(\eta) v_2(\tilde{\eta}) - v_1(\tilde \eta) 
v_2(\eta)
}{v_1'(\tilde \eta) v_2(\tilde \eta) - v_1(\tilde \eta) v_2'(\tilde \eta)}\, ,
\eeq
in terms of the two homogeneous solutions $v_1$ and $v_2$, which satisfy $\hat 
{\cal L} v_i = 0$, for a general differential operator $\hat {\cal L}$.
During the radiation dominated era 
the Green's function for the gravitational wave problem (\ref{equ:g}) reduces to
\beq
g''_k + k^2 g_k = \delta(\eta - \tilde \eta) \, ,
\eeq
which has the following homogeneous solutions
\beq
v_1 = \sin (k \eta)\, , \quad v_2 = \cos(k \eta)\, .
\eeq
Hence, the Green's function during the radiation dominated era is
\beq
g_k(\eta; \tilde \eta) = \frac{1}{k} [ \sin (k \eta) \cos (k \tilde \eta) - 
\sin(k \tilde \eta) \cos(k \eta)]\, , \quad \quad \eta < \eta_{\rm eq} \, .
\eeq
During matter domination 
equation (\ref{equ:g}) reduces to
\beq
g''_k + \Bigl( k^2 - \frac{2}{\eta^2} \Bigr) g_k = \delta(\eta - \tilde \eta)\, ,
\eeq
which has the following homogeneous solutions
\beq
v_1 = \eta\,  j_1(x)\, , \quad 
v_2 = \eta\, y_1(x)\, , \quad \quad x \equiv k \eta\, ,
\eeq
where $j_1(x)$ and $y_1(x)$ are spherical Bessel functions.
The Green's function during the matter dominated era therefore is
\beq
g_k(\eta; \tilde \eta)= - \frac{x \tilde x}{k}  \Bigl[j_1(x) y_1(\tilde x) - 
j_1(\tilde x) y_1(x) \Bigr]\, , \quad \quad \eta > \eta_{\rm eq}\, .
\eeq

\section{Transfer Function for First-Order Scalar Modes}
\label{sec:Bardeen}

\begin{figure}[h] 
   \centering
   \includegraphics[width=0.6\textwidth]{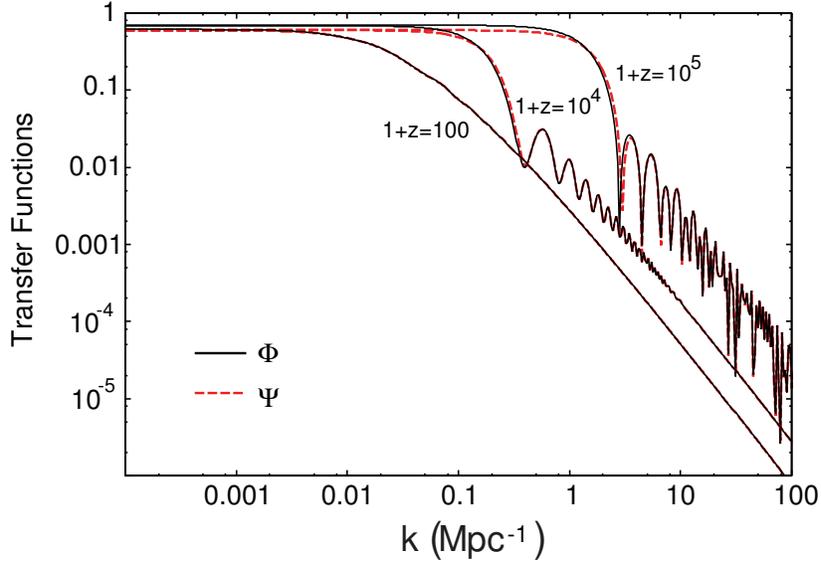} 
   \caption{{\bf Transfer functions for $\Phi$ and $\Psi$.} Neutrino anisotropic stress leads to ${\cal O}(10)$\% difference between $\Phi$ and $\Psi$ \cite{Hu:1995fq}.}
   \label{fig:trans}
\end{figure}

The first-order scalar perturbations $\Phi$ and $\Psi$ in equation (\ref{equ:metric}) satisfy the following constraint equation \cite{Mukhanov}
\beq
k^2 (\Phi-\Psi) = -4 \kappa^2 a^2 \left[ \rho_\gamma \Theta_2 + \rho_\nu {\cal 
N}_2 \right]\, ,
\eeq
where $\Theta_2$ and ${\cal N}_2$ characterize the quadrupole moments of the photon ($\gamma$) and neutrino ($\nu$) anisotropies, respectively. 
$\Theta_2$ and ${\cal N}_2$ are determine by the solution to the Einstein-Boltzmann equations.
In practice, these are solved numerically using CMBFAST \cite{CMBFAST} or CAMB \cite{CAMB} (see Figure \ref{fig:trans}). \\

Since, $\Theta_2$ and ${\cal N}_2$ are typically negligibly small, analytical studies often assume $\Phi \approx \Psi$.
In this case,
the first-order equation of motion for the Bardeen potential is ({\it e.g.} 
\cite{Mukhanov})
\beq
\Phi'' + 3(1+c_s^2) \cH \Phi' -c_s^2 \Delta \Phi + \Bigl( 2 \cH' + (1+3 c_s^2) 
\cH^2\Bigr) \propto \delta S\, ,
\eeq
where the RHS is non-zero only in the presence of entropy perturbations, $\delta S$.
For $p = w \rho$ and in the absence of entropy perturbations ({\it i.e.} $\delta 
S = 0$) this becomes
\beq
\label{equ:Bardeen}
\Phi_{\mathbf{k}}'' + \frac{6(1+w)}{1+3w} \frac{1}{\eta} \Phi_{\mathbf{k}}' + w 
k^2 \Phi_{\mathbf{k}} = 0\, .
\eeq
Equation (\ref{equ:Bardeen}) has the following exact solution
\beq
\Phi_{\mathbf{k}}(\eta) = y^{-\alpha} \Bigl[ C_1(k) J_\alpha(y) + C_2(k) 
Y_\alpha(y)\Bigr]\, , \quad \quad y \equiv \sqrt{w} k \eta\, , \quad  \alpha 
\equiv \frac{1}{2} \left(\frac{5+3 w}{1 + 3w} \right)\, ,
\eeq
where $J_\alpha$ and $Y_\alpha$ are Bessel functions of order $\alpha$.
During the matter dominated era ($w=0$) this becomes
\beq
\Phi_{\mathbf{k}}(\eta) = C_1(k) + \frac{C_2(k)}{y^5}\, ,
\eeq
whereas during the radiation dominated era ($w = \frac{1}{3}$) we find
\beq
\Phi_{\mathbf{k}}(\eta) = \frac{1}{y^2} \left[ C_1(k) \Bigl( \frac{\sin y}{y} - 
\cos y \Bigr) + C_2(k) \Bigl( \frac{\cos y}{y} + \sin y \Bigr) \right]\, .
\eeq
At early times $y = \sqrt{w} k \eta \ll 1$ this becomes asymptotically the 
primordial value,
\beq
\lim_{y \to 0} \Phi_{\mathbf{k}}(\eta) = C_1(k) = \psi_{\mathbf{k}}\, ,
\eeq
where we have dropped the decaying mode ($C_2 \equiv 0$).
For the growing mode solution we therefore obtain the following transfer function
\beq
\Phi(k \eta) = \left\{ 
\begin{array}{cr}
\frac{1}{(k \eta)^2} \Bigl(\frac{\sin [k \eta]}{k \eta} - \cos[k \eta] \Bigr) & 
\eta < \eta_{\rm eq}\\
const. &  \eta > \eta_{\rm eq}  
\end{array}
\right. 
\eeq
From this we see that
superhorizon modes ($k \eta \ll 1$) freeze  during the radiation era
\beq
\Phi(k \eta) = 1 + {\cal O}((k \eta)^2)\, , \quad k \eta \ll 1\, , \quad \eta < 
\eta_{\rm eq}\, ,
\eeq
while subhorizon modes ($k \eta > 1$) oscillate and decay as $a^{-2}$ 
\beq
\Phi(k \eta)
= - \frac{\cos[k \eta]}{(k \eta)^2}
= - \Bigl(\frac{\eta_k}{\eta} \Bigr)^2 \cos[k \eta]
= - \Bigl(\frac{a_k}{a} \Bigr)^2 \cos[k \eta] \, , \quad k \eta > 1\, , \quad 
\eta < \eta_{\rm eq}\, .
\eeq
Ignoring oscillations we therefore may write the following expression valid for both 
superhorizon and subhorizon modes
\beq
\label{equ:transfer}
\Phi(k \eta) = \frac{1}{1+ (k \eta)^2}\, , \quad \eta < \eta_{\rm eq}\, .
\eeq
The Bardeen potential freezes on all scales during matter domination.

\newpage
\begingroup\raggedright\endgroup

\end{document}